\documentstyle[aps,epsf,preprint,floats,tighten]{revtex}
\begin{document}
%\draft
\title{The origin of extra modes in
(VO)$_2$P$_2$O$_7$ }
\author{Kedar Damle$^{1,2}$ and S. E. Nagler$^3$}
\address{$^1$Department of Electrical Engineering, Princeton University,
Princeton, NJ 08544\\
$^2$Department of Physics, Princeton University, Princeton, NJ 08544\\
$^3$Oak Ridge National Laboratory, Oak Ridge, Tennessee 37831-6393}
\maketitle
\begin{abstract}
Recent inelastic neutron scattering experiments on (VO)$_2$P$_2$O$_7$ (VOPO)
have
seen previously unanticipated sharp peaks in the dynamic
structure factor in addition to the pair of triplet modes observed earlier.
We argue that the additional features are in essence
`shadows' of the previously studied features arising due
to umklapp scattering, and suggest
experimental tests of this proposal. The basic point is illustrated
by some elementary calculations within a strong-coupling 
expansion for  the alternating chain with the right geometry
(on which the matrix element for umklapp scattering
depends sensitively) taken fully into account. 
\end{abstract}

%\pacs{PACS numbers: 25.40.Fq, 75.10.Jm}

\section{Introduction}
\label{intro}
In recent years, there has been a great deal of experimental
and theoretical activity studying the magnetic properties of
insulating compounds that consist of arrays of well-isolated
one-dimensional magnetic sub-structures. An example is the compound
(VO)$_2$P$_2$O$_7$ (VOPO), which was initially thought to
be an excellent candidate for a spin-ladder compound based on
early experimental and theoretical work~\cite{eccle}.  Inelastic
neutron scattering experiments~\cite{oldexp} on single crystal arrays
established that these early ideas were incorrect and the
structure consists, to a good approximation, of an array
of alternating antiferromagnetic chains that are weakly coupled
to each other in one direction perpendicular to the chain axis.
The spectrum of magnetic excitations was mapped out by these
inelastic neutron scattering experiments. The lowest lying
excitation seen is a triplet mode separated by a gap from
the singlet ground state of the system. This is identified
with the basic single particle excitation expected theoretically
in an alternating chain (see for instance Ref~\cite{uhschu} and
references therein). The experiment also saw an
additional triplet mode above this band in a large
part of the Brillouin zone. The origin of the second mode was unclear, and
there
was some speculation that it could be
ascribed to a triplet bound state of the elementary excitations
that is also expected to exist in these systems~\cite{uhschu}. Such a state
may be stabilized 
by frustrating interactions~\cite{uhnorm}
It is more likely that the two
inequivalent magnetic
chains in VOPO have differing gap energies~\cite{kikuchi}; the second
mode would then be the basic triplet mode of the second set of chains.

New inelastic neutron scattering experiments on a single crystal of 
(VO)$_2$P$_2$O$_7$ (VOPO)~\cite{ens} have been able to map out 
the dispersion of the basic excitations of the system in greater
detail.
These experiments however also see {\em  additional sharp low-energy modes}
with dispersions different from the modes seen previously.
These extra modes are, at first sight, extremely surprising
and it is tempting to take them to be a signal of some new, and
hitherto unanticipated features in the spectrum of the  system (possibly
arising from frustrated couplings between alternating chains). However,
we argue that the real explanation for the new
modes is quite simple: both modes arise from a purely
geometric effect having to do with the actual positions of the
vanadium ions in the unit cell.  The
two new modes may be thought of as shadows of the basic triplet modes
arising from umklapp scattering.

We begin by detailing the geometry involved and use a very simple
general argument to calculate the matrix element for the
umklapp scattering process that is responsible for producing
a shadow of the basic
triplet mode. We then suggest a
straightforward check of this explanation based on a
comparison of the experimentally observed intensities of the
basic mode and its shadow at various values of the momentum
transfer. It is important to emphasize at this stage that
this check is quite independent of any theoretical estimates of
the intensity of the basic mode as a function of momentum
transfer and relies only on relations between experimentally
observed intensity ratios; the calculation
of the intensity of the basic single particle triplet mode as a function
of $k$ ($2 \pi k/b \equiv k_b$ is the momentum transfer along
the chain direction, where $b$ is the unit cell dimension
along the chain axis, which is conventionally
labeled the b axis) is a separate problem that has been
addressed earlier for the simple alternating
chain~\cite{barnes} (these results
may also be used in conjunction
with our analysis to give approximate intensities of the shadow of the
single particle band, but we do not perform that
exercise here). A simple consequence of this
scenario is the prediction that the shadow band will {\em disappear for}
$h=0$ (here $2\pi h/a \equiv k_a$ is the momentum transfer along the
crystallographic
$a$
axis perpendicular to the  alternating chain axis).

One also expects
that shadows of any bound-state mode will also be formed
by an analogous mechanism involving umklapp scattering.
We illustrate this by an explicit calculation, to leading
order in a strong-coupling expansion, of 
the bound state contribution to the dynamic structure factor for
an alternating chain with the right geometry taken into account.
The calculated intensity ratios do provide an explicit example of the
general argument for the strength of the umklapp contribution.

We also briefly explore
the possibility that the magnetic interactions felt by even and
odd dimers (pairs of spins connected by the
stronger of the two antiferromagnetic interactions in an
alternating chain model) are slightly different. We expect that
this will change the strength of the shadow bands in a
significant way.
To get a feel for what to expect, we do a simple calculation, again within a 
strong
coupling expansion, of the  contribution of the basic
triplet mode to the dynamic structure factor for an alternating chain
with the right geometry and the small difference in magnetic
interactions felt by even and odd dimers. We see that this change
in the magnetic environments of even and odd dimers
leads to a weak intensity for the shadow mode even
at $h=0$ (in contrast to our
result for the simpler alternating chain of Fig~\ref{figchain1})
as well as a small splitting between the basic mode and
its shadow at $k=1/2 \, , \, 3/2$ in the fundamental Brillouin zone at
$h=0$. This shadow
at $h=0$, as well as the splitting at $k=1/2 \, , \, 3/2$
are a sensitive test of the difference in the magnetic environments
of even and odd dimers in the chain.
All experiments to date~\cite{oldexp,ens,agthesis} are
consistent with the absence of a shadow mode at $h=0$. However, in
the absence of any straightforward symmetry reason forcing the
even and odd dimers to be equivalent, the
possibility that more refined experiments with better
statistics will see a weak shadow is still open.

Lastly, it must be emphasized that our entire approach here ignores the
frustrating couplings between chains that has been argued to
exist based on the small, but experimentally detectable dispersion
seen as a function of $h$.  These couplings
along the $a$ direction are important ingredients
of any quantitatively accurate calculation of the expected
neutron scattering intensity, but are not expected to change
significantly any of our conclusions regarding the shadow modes.

\section{Shadow bands due to umklapp scattering.}
\label{shadow}
We begin with a brief review of some of the relevant
details of the structure~\cite{nguyen} of VOPO:
Previous work~\cite{oldexp} on VOPO has established that
the compound may be thought of as an array of alternating
antiferromagnetic
chains (with the chains oriented along the crystallographic $b$
axis and the V$^{4+}$ ions forming the basic spin $1/2$
constituents of the chains) that are weakly coupled to each other in the $a$
direction, and essentially decoupled in the $c$ direction.
As mentioned in the introduction, we will, for the most part,
ignore the weak interchain coupling as it is not
expected to materially change any of our conclusions.\begin{figure}
\epsfxsize=5.0in
\centerline{\epsffile{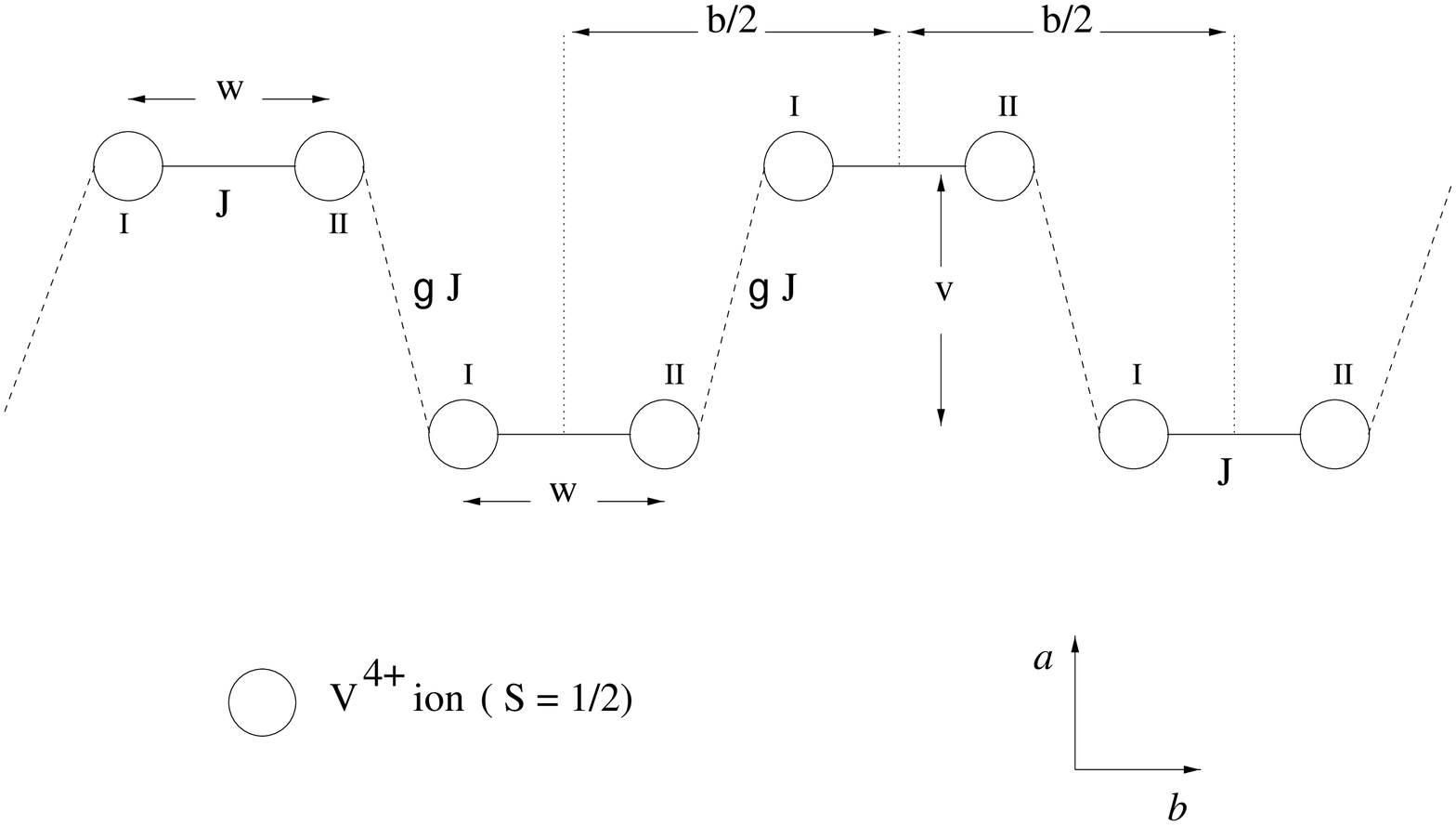}}
\vspace{0.1in}
\caption{Geometry of the alternating chain. Note that the
staggering of the dimers in the $a$ direction is greatly
exaggerated in the figure; $v \approx 0.09 a$, while $w \approx 0.31 b$
with $a \approx 7.7$ angstroms and $b \approx 16.6$ angstroms.}
\label{figchain1}
\end{figure}
Each unit cell of VOPO contains eight V$^{4+}$ ions, comprising four dimers.
Each of the dimers belongs to a different chain. There is a pair of chains
near $z=0$,
and a second pair near $z=c/2$. The members of each pair are related to
each other via
a screw axis transformation and therefore there are at most two magnetically
distinct chains displaced from each other along the $c$ axis.
We can thus focus attention on one representative from each pair. 
These two chains have similar (but not identical) 
structures and the geometrical
effects we discuss here are very nearly identical for each type of chain.
Using the known structure, we
can draw out the actual positions of the
vanadium ions in the $a-b$ plane. 
To a good approximation, this gives us
an alternating chain along the $b$ axis in which
successive dimers are slightly staggered along
the $a$ axis as shown in Fig~\ref{figchain1}.
There is an extremely tiny displacement
of the ions relative to each other in the $c$ direction; this
is small enough that we feel justified in ignoring it in our
analysis. Similarly we can ignore tilts of the dimer units away from 
the $b$ axis.

The magnetic response of a single
chain may be modeled (modulo the possible
complications that form the subject matter
of section~\ref{magnetic}) by the simple alternating
chain Hamiltonian
\begin{equation}
{\cal H} = J \sum \limits_{i}[{\bf S}_{I}(i) \cdot {\bf S}_{II}(i)
+g{\bf S}_{II}(i) \cdot {\bf S}_{I}(i+1)] \; ,
\label{simpleH}
\end{equation}
where $J$ is the overall energy scale fixed by the microscopic
exchange constants in the system, $g$ represents the
ratio of the weak and the strong bonds of the alternating chain,
and the spins are labelled as in Fig~\ref{figchain1}. Note that
this Hamiltonian is invariant under translations of
$b/2$ along the $b$ axis. However,
a glance at Fig~\ref{figchain1} shows that
the staggering of the
positions of even and odd dimers in the $a$ direction
reduces the  actual symmetry of the full structure to
translations by $b$ and not $b/2$ along the $b$ axis.
This of course implies that momentum conservation may
be violated during a neutron scattering event by
integer multiples of $2\pi/b$ along the chain axis. Note that
this is less stringent than the more usual
condition (which would be in force in the absence of any staggering
along the $a$ axis of even and odd dimers) 
that momentum be conserved modulo integer multiples of
$4\pi/b$, and we believe that this simple fact is at the
root of the observed shadow bands. Thus, we expect that the extra modes
seen should be displaced by precisely $2\pi/b$ from the basic modes
of the alternating chain. This seems to be the case with the
experimental data~\cite{ens}. To clinch the identification,
we need to be able to make predictions for the intensities of the 
extra modes relative to the basic modes and see how these compare
with the experimental numbers for the intensity ratios. This is what
we turn to next.

Let us begin our analysis by 
writing down the usual spectral
representation for the dynamic structure factor of our system
at $T=0$:
\begin{eqnarray}
S_{zz}({\bf k}, \omega) & = & \sum \limits_{N} \delta(\omega - 
E_N + E_0)\left |\langle \Phi_N| S^{z}(-{\bf k})|\Phi_0\rangle \right|^2
\; , 
\end{eqnarray}
where $|\Phi_0\rangle$ is the exact ground state of the
system, $|\Phi_N \rangle$ is an exact excited state labeled
by the index $N$, $E_0$ and $E_N$ are the energies of the
ground and the excited state respectively and $S^{z}(-{\bf k})$ is
defined as
%
% indices below changed to j
%
%
\begin{equation}
S^{z}(-{\bf k}) = (\frac{b}{4L})^{1/2} \sum \limits_{j\,A} S^{z}_{A}(j)
e^{i{\bf k} \cdot {\bf x}_{j\, A}} \; ,
\end{equation}
where the subscript $A$ takes on values $I$ and $II$, $j$ refers to
the dimer index, $L$ is the length of
the chain and ${\bf x}_{j \, A}$ is the position of the
spin labeled by $j$ and $A$ (see Fig~\ref{figchain1}) (here and in
the rest of our discussion, we will exploit the rotational
invariance in spin space to focus only on the $zz$ component
of the dynamic structure factor).

It is convenient to formulate our analysis
in terms of operators that directly make reference to the
states of each strongly coupled dimer in the system. This
is achieved by transforming to the so-called `dimer boson'
representation~\cite{leuen,ssbhatt,cowley}.
Following Ref~\cite{ssbhatt}, we write
the spin operators as:
%
% indices below changed to j
%
%
\begin{eqnarray}
S_{I}^{\alpha}(j) & = & \frac{1}{2}\left(s^{\dagger}(j)t_{\alpha}(j) +
t_{\alpha }^{\dagger}(j)s(j)-i\epsilon_{\alpha \beta
\gamma}t_{\beta}^{\dagger}(j)
t_{\gamma}(j)\right )~, \label{bonddefI} \\
S_{II}^{\alpha}(j) & = & \frac{1}{2}\left(-s^{\dagger}(j)t_{\alpha}(j) -
t_{\alpha }^{\dagger}(j)s(j)-i\epsilon_{\alpha \beta
\gamma}t_{\beta}^{\dagger}(j)
t_{\gamma}(j)\right )~,
\label{bonddefII}
\end{eqnarray}
where $\alpha$, $\beta$, and $\gamma$ are vector indices taking the values
$x$,$y$,$z$, repeated indices are summed over, and $\epsilon$ is the totally
antisymmetric tensor. $s^{\dagger}(j)$ and $t_{\alpha}^{\dagger}(j)$ are
respectively creation operators for singlet and
triplet bosons at `site' $j$ (in the bosonic language, each strongly
coupled dimer is thought of as a single site; the separation of
adjacent sites along the $b$ axis is then $b/2$).
The restriction that physical states of
a dimer are either singlets or triplets leads to the following constraint on
the boson occupation numbers at each site:
%
%indices changed to j
%
\begin{displaymath}
s^{\dagger}(j)s(j)+t_{\alpha}^{\dagger}(j)t_{\alpha}(j)=1~.
\end{displaymath}
The spin density is given by
\begin{displaymath}
\sigma_{\alpha}(j) = -i\epsilon_{\alpha \beta \gamma}t_{\beta}^{\dagger}(j)
t_{\gamma}(j)~.
\end{displaymath}
It is also convenient to define
\begin{displaymath}
\phi_{\alpha}(j)=s^{\dagger}(j)t_{\alpha}(j) +t_{\alpha }^{\dagger}(j)s(j)~.
\end{displaymath}
Note that as the constraint fixes the number of singlet particles
uniquely given the triplet occupation number, we may as
well think only in terms of triplet occupation numbers; we will
thus refer to any site which is occupied by a singlet as being
in the vacuum state.

Finally, it is useful to note that the alternating chain is 
readily analyzed in the
limit of strong alternation (the so called strong coupling
limit, with $g \ll 1$)~\cite{uhschu,barnes}. In the language
we are using here, the lowest lying excitations in this limit
are single particle modes (with one bare triplet particle
excited above the ground state, which may be thought of
as vacuum). While 
corrections are certainly introduced to this picture at
higher orders in $g$, it is still legitimate
to think of the basic triplet mode seen in the real system
as arising from the contribution of the fully renormalized single
particle excitation in the system (for a careful analysis
of this point for the closely related problem
of a spin-ladder, see Ref~\cite{sskd})

With these preliminaries out of the way, let us now go through the
extremely elementary general argument for the
strength of the umklapp matrix element responsible for the
shadow bands. We formulate this here only for contributions
to the spectral sum coming from the fully renormalized single
particle states of the system. The basic argument is nevertheless
expected to remain valid when applied to the contributions coming
from two-particle bound states; we will see this expectation
realized in an explicit calculation later.

Let us begin by writing the contribution of the fully renormalized
single particle states as
\begin{equation}
S_{zz}^{{\rm{1p}}}({\bf k}, \omega) = \sum \limits_{q = 0}^{4\pi/b}
\delta(\omega - \varepsilon_q)
\left| \langle q| S^{z}(-{\bf k}) |\Phi_0\rangle\right|^2 \; ,
\label{qsum1}
\end{equation}
where $|q\rangle$ is the exact, fully renormalized single particle
state  of momentum $q$ in the chain direction (the state
of course has $z$ component of its spin equal to $0$; we will
not be very careful in this section about including this information
in our notation) and $\varepsilon_q$
is the energy of this state (with the ground state energy
set to zero). We may write this
state quite generally as
\begin{equation}
|q\rangle = (\frac{b}{2L})^{1/2} \sum \limits_{j}e^{iqx_j^{b}} | \Psi_1(j)
\rangle \; ,
\end{equation}
where $x_j^{b}$ is the $b$ component of the position vector of the
$j^{th}$ `site' (center of the $j^{th}$ dimer) in the chain  and the
notation $|\Psi_1(j)\rangle$ is intended to denote a state that
differs from $|\Phi_0\rangle$ only locally in the vicinity of
site $j$. Furthermore, we can write $S^{z}(-{\bf k})$ as
\begin{equation}
S^{z}(-{\bf k}) = 
(\frac{b}{2L})^{1/2}
(
e^{ik_a v/2} \sum \limits_{j \, {\rm odd}}e^{ik_bx_j^{b}} {\cal O}_{k_b}(j)
+ e^{-ik_a v/2} \sum \limits_{j \, {\rm even}}e^{ik_bx_j^{b}} {\cal
O}_{k_b}(j)
)  \; ,
\label{szo}
\end{equation}
where the operator ${\cal O}_{k_b}(j)$ is defined as
\begin{equation}
{\cal O}_{k_b}(j) = \frac{1}{\sqrt{2}}(\cos(\frac{ k_bw}{2}) \sigma_{z}(j)
- i \sin(\frac{k_bw}{2}) \phi_{z}(j)) \; ,
\label{o12}
\end{equation}
with $w$ equal to the distance along the $b$ axis between the two
spins that form each dimer.

This now allows us to write the following expression for the
matrix element appearing in the spectral sum (\ref{qsum1}):
\begin{equation}
\left| \langle q| S^{z}(-{\bf k}) |\Phi_0\rangle\right|^2
=\frac{|{\cal M}(q,k_b)|^2}{4} |e^{-ik_av/2} + e^{i(k_b-q)b/2}e^{ik_av/2}|^2 
{\tilde \delta}_{q \, , \, k_b} \; ,
\label{me1}
\end{equation}
where ${\tilde \delta}$ is defined as
\begin{equation}
{\tilde \delta}_{q \, , \, k_b} = \sum \limits_{n = - \infty}^{\infty}
\delta_{q \, , \, k_b + 2\pi n/b} \; ,
\end{equation}
and ${\cal M}$ is given as
\begin{eqnarray}
{\cal M}(q,k_b) & = & \sum \limits_{j} e^{-iqx_{j}^{b}}\langle \Psi_1(j)|{\cal
O}_{
k_b}(0)|\Phi_0\rangle \; . 
\end{eqnarray}
Note that we can make the $k_b$ dependence of ${\cal M}$ explicit
by rewriting this as
\begin{equation}
{\cal M}(q,k_b) = \cos(\frac{ k_bw}{2}){\cal M}_{\sigma}(q) +
\sin(\frac{ k_bw}{2}){\cal M}_{\phi}(q) \; ,
\end{equation}
where
\begin{eqnarray}
{\cal M}_{\sigma}(q) &=& \frac{1}{\sqrt{2}}
\sum \limits_{j} e^{-iqx_{j}^{b}}\langle \Psi_1(j)|\sigma_{z}(0)|\Phi_0\rangle
\; , \nonumber \\
{\cal M}_{\phi}(q) & = & \frac{-i}{\sqrt{2}}
\sum \limits_{j} e^{-iqx_{j}^{b}}\langle \Psi_1(j)|\phi_{z}(0)|\Phi_0\rangle
\; .
\end{eqnarray}
Furthermore, previous work~\cite{barnes} has demonstrated that 
${\cal M}_{\sigma}$
is identically zero for the single particle states of the
simple alternating chain Hamiltonian~(\ref{simpleH}).

Using all of this we can write the contribution of the fundamental
single-particle mode to the dynamic structure factor as
%
% left the ka kb as is.
%
\begin{eqnarray}
S_{zz}^{{\rm{1p, basic}}}({\bf k}, \omega) & = & |{\cal M}_{\phi}(k_b)|^2 
\sin^2(\frac{k_b w}{2}) \cos^2(\frac{k_a v}{2}) \delta(\omega - 
\varepsilon_{k_b}) \; \; \; k_b \, \in \, (0,\frac{4\pi}{b})\;, \nonumber \\
& = &  |{\cal M}_{\phi}(k_b-4\pi/b)|^2 
\sin^2(\frac{k_b w}{2}) \cos^2(\frac{k_a v}{2}) \delta(\omega - 
\varepsilon_{k_b - 4\pi/b}) \; \; \; k_b \, \in \, 
(\frac{4\pi}{b},\frac{8\pi}{b})\;,
\end{eqnarray}
while the shadow contribution reads
\begin{eqnarray}
S_{zz}^{{\rm{1p, shadow}}}({\bf k}, \omega) & = & 
|{\cal M}_{\phi}(k_b+2\pi/b)|^2 
\sin^2(\frac{k_b w}{2}) \sin^2(\frac{k_a v}{2}) \delta(\omega - 
\varepsilon_{k_b+2\pi/b}) \; \; \; k_b \, \in \, 
(0,\frac{2\pi}{b})\;, \nonumber \\
& = & 
|{\cal M}_{\phi}(k_b-2\pi/b)|^2 
\sin^2(\frac{k_b w}{2}) \sin^2(\frac{k_a v}{2}) \delta(\omega - 
\varepsilon_{k_b-2\pi/b}) \; \; \; k_b \, \in \, 
(\frac{2\pi}{b},\frac{6\pi}{b})\;, \nonumber \\
& = & 
|{\cal M}_{\phi}(k_b-6\pi/b)|^2 
\sin^2(\frac{k_b w}{2}) \sin^2(\frac{k_a v}{2}) \delta(\omega - 
\varepsilon_{k_b-6\pi/b}) \; \; \; k_b \, \in \, 
(\frac{6\pi}{b},\frac{8\pi}{b})\;.\nonumber \\
&&
\label{shadrealcomp1}
\end{eqnarray}

Thus, we see quite generally that the intensity of the
shadow band should vanish as $k_a \rightarrow 0$ (modulo
the complications discussed in section~\ref{magnetic}). Moreover,
it is apparent from these expressions that the intensity of
the shadow is completely determined by the intensity of
the basic mode as a function of $k_b$. While there are, in
principle, a number of ways in which this may be checked against
the experimental data of Ref~\cite{ens}, it is probably best to
simply use the experimentally observed intensity of the fundamental
mode at fixed $k_a$ (for values of $k_b$ at which the `dimer coherence factor'
$\sin^2(k_bw/2)$ is large) to predict the intensity of
the shadow {\em at the same $k_a$} (this avoids complications
due to the weak two-dimensional couplings between chains that will
introduce additional dependence of the intensity on $k_a$). This prediction
can then be directly tested against the observed intensity of the
shadow after correcting for effects of the magnetic form factor
of the V$^{4+}$ ion. Note that this procedure makes no assumptions about
the form of ${\cal M}_{\phi}(k_b)$ for the alternating chain Hamiltonian
(\ref{simpleH}) and serves to separate the purely geometric
effect leading to the shadow from our approximate knowledge
of this function.

Let us conclude this section by noting that entirely analogous
arguments can be used to relate the expected intensity of the
shadow of a bound-state mode to the intensity of the
bound-state itself (of course, the analog of
${\cal M}_{\sigma}$ is no longer identically zero, but this
merely complicates the algebra a little). Instead of going through
the corresponding argument for the bound state modes in detail, we
choose to highlight the minor differences
involved by doing an approximate calculation of the intensity and
position of both the bound-state mode and its shadow to
leading order in a strong-coupling expansion. This is what we turn to
in the next section.

\section{Bound-state contributions within the strong-coupling expansion}
\label{bssc}
The first order of business is to work out the position in
the Brillouin zone and the energy of the $S=1$ bound state formed
from the physical (fully renormalized) triplet particles that
are the elementary excitations of the alternating chain (\ref{simpleH}).
To leading order in the strong coupling expansion, this is
particularly simple as the physical single-particle excitation
coincides with the bare triplet particle created by the triplet
boson operator as far as the energy levels are concerned. Following
the approach used in Ref~\cite{sskd}, it is easy to see~\cite{kdup} at
leading order that
the triplet bound state exists over two separate
intervals for the center of mass momentum $q_{cm}$
%
% didnt define qcm explicitly as presumably it has
%standard meaning.
%
(note that
the center of mass momentum takes on values in the range $(0, 8\pi/b)$):
the first being $(4\pi/3b, 2\pi/b)$ and the second being $(6\pi/b,20\pi/3b)$.
The energy of the bound state (with the ground state energy set to
zero) is given to leading order as $\varepsilon_B(q_{cm})/J =
2 - g(4\cos^2(q_{cm}b/4)+1)/4$ (these results were first
obtained for a slightly more general Hamiltonian by Uhrig and 
Schulz~\cite{uhschu}).

The next thing we need is the bound state wavefunction and the
ground state wavefunction correct to first order in $g$; note that it
does not suffice to know these eigenstates to leading (zeroth) order
in $g$ as it turns out that the bound state contributes to the
dynamic structure factor only at first order or higher in $g$.
We will write these eigenstates down in the basis of (bare) triplet
boson occupation numbers and polarizations. An extremely
elementary
calculation~\cite{kdup} gives us the following ground state, correct
to first order in $g$:
\begin{equation}
|\Phi_0\rangle = |0\rangle + \frac{g}{8}\sum \limits_{j}\left(
|(j)[0],(j+1)[0]\rangle -|(j)[-1],(j+1)[+1]\rangle-|(j)[+1],(j+1)[-1]\rangle
\right)\; ,
\end{equation}
where $|0\rangle$ represents the vacuum state for the
triplet bosons,  the number in the square brackets gives the $z$ component
of the spin of the triplet boson and the number in
the parenthesis gives the site occupied by the boson (two such
pairs separated by a comma naturally denote a two-particle state
in the bosonic Fock space).

The zeroth order normalized bound state labeled by the
center-of-mass momentum $q_{cm}$ (and $z$ component of
spin equal to $0$) can be easily calculated~\cite{kdup} to
be 
\begin{equation}
|q_{cm}[0]\rangle = \sum \limits_{j_2 > j_1}f_{q_{cm}}(j_1,j_2)
\left( |(j_1)[-1],(j_2)[+1]\rangle - |(j_1)[+1],(j_2)[-1]\rangle \right) \; ,
\end{equation}
where the bound-state wavefunction $f_{q_{cm}}$ is given as
\begin{equation}
f_{q_{cm}}(j_1,j_2) = (\frac{b}{2L})^{1/2}(\frac{e^{\kappa b}-1}{2})^{1/2}
e^{-\kappa|x^b_{j_2} - x^b_{j_1}|}e^{iq_{cm}(x^b_{j_2}+x^b_{j_1})/2} \; ,
\end{equation}
with $e^{-\kappa b/2} = 2 \cos(q_{cm}b/4)$.

Now, following the approach used in Ref~\cite{sskd}, it is quite
easy to see that the ${\cal O}(g)$ corrections to this will
involve states living in the zero, one, three and four boson
sectors of the Fock-space for the
bare triplet bosons,  in addition to a ${\cal O}(g)$ correction to the
component in the two-boson sector. It is an elementary
exercise~\cite{kdup} to work out all these corrections except the
one in the two-boson sector (as this involves first working
out the effective dynamics of the physical particles to one higher
order in $g$). While the correction in the two-boson sector can
also be calculated without too much difficulty using the methods
referred to earlier, we do not bother to do this explicitly here as
it is quite clear that this correction term  plays no role
in the leading order calculation of the bound state contribution to
the dynamic structure factor. In fact, for our purposes here,
it clearly suffices to work out the correction in the one boson
sector of the Fock space as this is the only correction that
affects our calculation.
This may be written down readily~\cite{kdup} as:
\begin{equation}
\delta^{(1)}|q_{cm}[0]\rangle = -\frac{g}{4}\sum \limits_{j}f_{q_{cm}}(j,j+1)
\left( |(j)[0]\rangle + |(j+1)[0]\rangle \right)\;,
\end{equation}
where $\delta^{(1)}|q_{cm}[0]\rangle$ is the first order correction
term in the one-boson sector.

With all this in place, we can begin our analysis of the
bound state contribution to the dynamic structure factor
by writing down the following spectral sum:
\begin{equation}
S^{\rm bs}_{zz}({\bf k},\omega)=\sum \limits_{q_{cm} = 0}^{8\pi/b} 
\delta(\omega - \varepsilon_B(q_{cm}))
\left| \langle q_{cm}[0]| S^{z}(-{\bf k}) |\Phi_0\rangle\right|^2 \; .
\label{qcmsum1}
\end{equation}
Notice the different range of summation for $q_{cm}$ in comparison
with Eqn~(\ref{qsum1}) (it is of course understood that the sum is performed 
only over those sub-intervals
in $q_{cm}$ that actually support the existence of a $S=1$
bound state).
We can now use (\ref{szo}) and (\ref{o12}) and calculate the
matrix element appearing in the spectral sum to first order in
$g$ using the ${\cal O}(g)$ wavefunctions calculated above.
 
While it is certainly possible to use the notation of section~\ref{shadow}
and
only quote the perturbative results for the analogs of
${\cal M}_{\sigma}$ and ${\cal M}_{\phi}$, we prefer to
put everything together and directly present results for
the leading contribution to the dynamic
structure factor.
%\begin{figure}            
%\epsfxsize=1.5in
%\centerline{\epsffile{figbs1.eps}}
%\vspace{0.1in}
%\caption{Position of the bound state mode and its shadow to leading
%order in the strong coupling expansion}
%\label{figbs1}
%\end{figure}
Naturally,
these results are not expected to be quantitatively accurate. Rather,
they provide us with a non-trivial example of the general
argument of section~\ref{shadow} at work.\begin{figure}
\epsfxsize=4.5in
\centerline{\epsffile{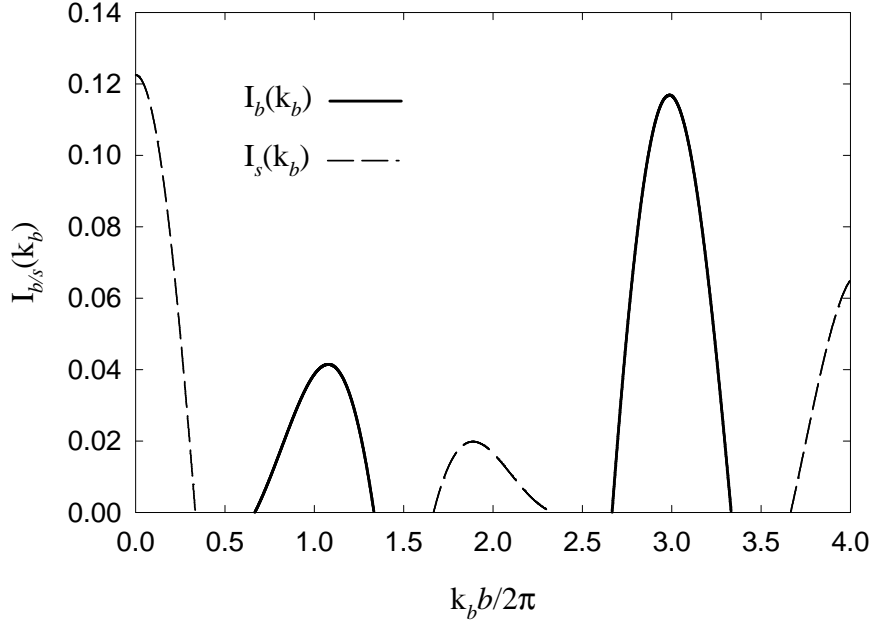}}
\vspace{0.1in}
\caption{Intensity of the bound state mode $I_b$, and the intensity
of the shadow $I_s$ to leading
order in the strong coupling expansion. Note that the value of
$k_a$ is different for the two; in each case it is chosen to
maximize the intensity. The intensities are both normalized by
the average value of the intensity in the single-particle mode
at this order in the strong coupling expansion. The value
of $g$ used is $0.7$, which is approximately right for VOPO~{\protect
\cite{oldexp}}}
\label{figbs2}
\end{figure}
The bound state leads
to the following basic contribution to the dynamic structure factor
for $k_b \in (4\pi/3b,8\pi/3b)$ and again for $k_b \in (16\pi/3b,20\pi/3b)$:
\begin{eqnarray}
S_{zz}^{\rm bs, basic}({\bf k},\omega) & = & 
\frac{g^2}{16}(1-4\cos^2(\frac{k_b b}{4}))\sin^2(\frac{k_b}{2}(w-\frac{b}{2}))
\cos^2(\frac{k_a v}{2}) \delta(\omega-\varepsilon_B(k_b))\;.
\end{eqnarray}
The shadow of the bound state gives for $k_b \in (0,2\pi/3b)$,
$k_b \in (10\pi/3b,14\pi/3b)$ and $k_b \in (22\pi/3b,8\pi/b)$:
\begin{eqnarray}
S_{zz}^{\rm bs, shadow}({\bf k},\omega) & = & 
\frac{g^2}{16}(1-4\sin^2(\frac{k_b b}{4}))\cos^2(\frac{k_b}{2}(w-\frac{b}{2}))
\sin^2(\frac{k_a v}{2}) \delta(\omega-\varepsilon_B^{\rm s}(k_b))\;,
\end{eqnarray}
where $\varepsilon_b^{\rm s}(k_b)/J =2 - g(4\sin^2(k_b b/4)+1)/4$ gives
us the position of the shadow band. 
The intensity in the bound state mode and in its
shadow is depicted are in Fig~\ref{figbs2}.
%and Fig~\ref{figbs1}.

We thus see that the bound state mode also acquires a `shadow'
as anticipated earlier on the basis of the general argument.
Of course, as mentioned previously, this strong-coupling
calculation has very little quantitative significance. A
quantitative analysis of any experimental data on the bound
state mode and it's shadow would instead follow the
analog of the procedure outlined the end of section~\ref{shadow}
with the analog of ${\cal M}_{\sigma}$ included in the
analysis and the necessary changes made to allow for the
fact that the spectral sum is to be carried out
over a different range from the single particle case (note that
this type of analysis can, in principle, distinguish between
bound state and single particle triplet modes based on the
different intensity ratios between the modes and their shadows in
the two cases).

\section{A more complicated magnetic Hamiltonian?}
\label{magnetic}
In this final section we briefly consider a
possible complication that will affect our results at a 
qualitative level:
namely the possibility that the magnetic interactions
felt by the even and the odd dimers are slightly different. This will
clearly change the intensities and the dispersions of the various
modes observed as the magnetic Hamiltonian will now be invariant
under translations of $b$ and not $b/2$ along the $b$ axis. Thus,
the staggering of the dimer positions along the $a$ axis will
no longer be the only thing breaking the larger symmetry of
translations by $b/2$. Clearly, in such a
situation, we expect that the `shadow'
band intensity will be non-zero even at $k_a = 0$ (indeed we
expect that it will depend quite sensitively on the
difference in the magnetic interactions of the even and
odd dimers). To get a feel for what to expect, let us
work out the intensity of the basic single particle
mode and its shadow to leading order within a
strong coupling expansion.

The Hamiltonian we have in mind can be parameterized as:
\begin{eqnarray}
{\cal H} &= &J \sum \limits_{j\, odd}[(1+g\lambda){\bf S}_{I}(j) 
\cdot {\bf S}_{II}(j)
+g(1+\mu){\bf S}_{II}(j) \cdot {\bf S}_{I}(j+1)] \nonumber \\
&& + \;
J \sum \limits_{j\, even}[(1-g\lambda){\bf S}_{I}(j) 
\cdot {\bf S}_{II}(j)
+g(1-\mu){\bf S}_{II}(j) \cdot {\bf S}_{I}(j+1)] \; , 
\label{compH}
\end{eqnarray}
where we have in mind that $\lambda$ and $\mu$ are both small
parameters that model the small differences in the 
magnetic properties of the even and the odd dimers.

The calculation of the ${\cal O}(g)$ single-particle contribution to the
dynamic structure factor is quite 
elementary and involves nothing
new. We will therefore be correspondingly brief.

We begin our analysis by noting that it is now more natural to
count states somewhat differently. We will restrict the
momentum carried by the single particle state to lie in the
range $(0, 2\pi/b)$ and allow for two distinct bands of
single particle states labeled by $+$ and $-$ subscripts.

The energies of these two bands are easily worked out to leading
order to
be
\begin{equation}
\varepsilon_{\pm}(q) = 1 \mp \frac{g}{2}(4 {\lambda}^2 + \cos^2(qb/2)
+{\mu}^2\sin^2(qb/2))^{1/2} \; .
\end{equation}
Moreover, it is quite elementary to see that the corresponding 
eigenstates
to leading order in the strong coupling expansion are (again we choose
to write down the state with $S_z = 0$ as this is what we
need to calculate the $zz$ component of the dynamic structure factor):
\begin{equation}
|q_{\pm}[0]\rangle =(\frac{b}{2L})^{1/2}\sqrt{2}{\cal P}(q)( \sum
\limits_{j \, odd} e^{iqx^b_{j}}|j[0]\rangle
+y_{\pm}(q) \sum \limits_{j \, even} e^{iqx^b_{j}}|j[0]\rangle) \; ,
\end{equation}
where ${\cal P}(q) = 1/\sqrt{1+|y_{\pm}(q)|^2}$ and
$y_{\pm}(q)$ is given as
\begin{equation}
y_{\pm}(q) = \frac{2\lambda \pm (4 {\lambda}^2 + \cos^2(qb/2)
+{\mu}^2\sin^2(qb/2))^{1/2}}{\cos(qb/2) + i \mu \sin(qb/2)}
\end{equation}

One other thing we need is the 
ground state for this model, correct to ${\cal O}(g)$. 
This can also be worked out
quite easily to be~\cite{kdup}
\begin{eqnarray}
|\Phi_0\rangle & = &|0\rangle + \frac{g(1+\mu)}{8}\sum \limits_{j \,
odd}\left(
|(j)[0],(j+1)[0]\rangle -|(j)[-1],(j+1)[+1]\rangle-|(j)[+1],(j+1)[-1]\rangle
\right) \nonumber \\
&& + \; \frac{g(1-\mu)}{8}\sum \limits_{j \, even}\left(
|(j)[0],(j+1)[0]\rangle -|(j)[-1],(j+1)[+1]\rangle-|(j)[+1],(j+1)[-1]\rangle
\right) \nonumber \\
&&
\end{eqnarray}

We can now use all of this to work out the one particle
piece of the dynamic structure factor.\begin{figure}
\epsfxsize=4.5in
\centerline{\epsffile{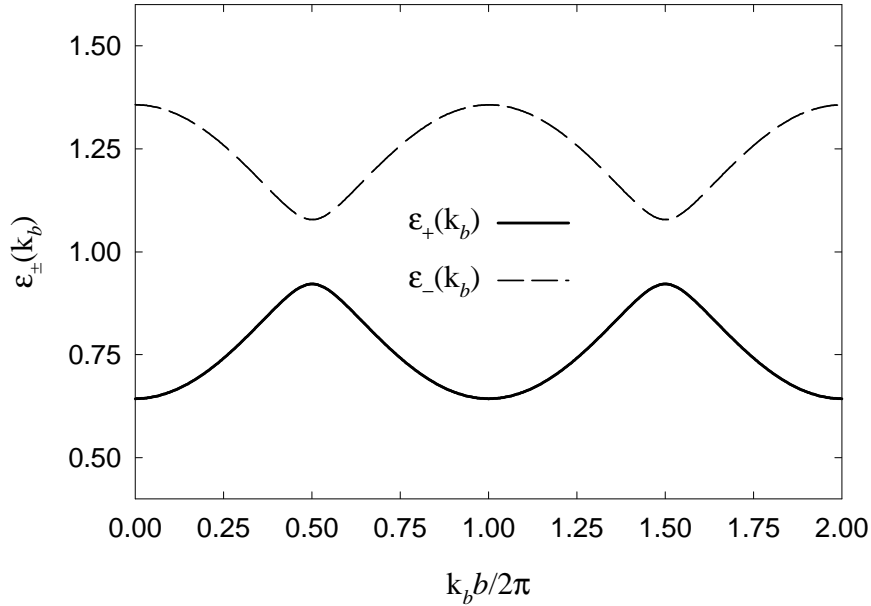}}
\vspace{0.1in}
\caption{Position of the two single particle
bands when the interactions felt by the even and
odd dimers are different. As explained in the text, the
`basic mode' and it's `shadow' are actually both hybrids
made up of the two single particle bands in the problem.
We have chosen, for illustrative purposes, the values $\lambda = 0.1$
and $\mu = 0.1$. The parameter $g$ is set equal to $0.7$, which
is approximately correct for VOPO~{\protect \cite{oldexp}}.}
\label{figmag1}
\end{figure}
The resulting expressions
are quite messy for general ${\bf k}$ and not particularly illuminating.
We will
write them down here only for the special case of $k_a =0$, as this
is where we expect a real qualitative difference due to the
complications we have introduced into the problem:
\begin{eqnarray}
S_{zz}({\bf k}, \omega) &=& \sum \limits*12 
_{\alpha =
\pm}\frac{\sin^2(k_bw/2)}{4}
\left[(1+\frac{2G_{\alpha}\cos(k_b b/2)}{G^2_{\alpha}+Q^2})
+ \frac{g}{2}(\cos(k_bb/2) + \frac{2Q^2G_{\alpha}}{G^2_{\alpha}+Q^2})\right]
\nonumber \\
&&~~~~~~~~~~~~~~~~~~~~~~\times\; \delta(\omega-\varepsilon_{\alpha}(k_b)) \; ,
\label{messy1}
\end{eqnarray}
where we have defined $Q^2$ as
\begin{equation}
Q^2(k_b) = \cos^2(k_b b/2) + \mu^2 \sin^2(k_b b/2) \; ,
\end{equation}
and $G$ as
\begin{equation}
G_{\pm}(k_b) = 2 \lambda \pm \sqrt{4\lambda^2 + Q^2(k_b)} \;.
\end{equation}
The details of the above expressions are not particularly important.
We only wish to use the above to arrive at some general qualitative
conclusions about the nature of the expected intensity at
various points in the Brillouin zone. The first of these is of course
that we have some non-zero intensity at the shadow positions even
at $k_a =0$. In this context, it is important to note that
both the `basic mode' and the `shadow' are actually
hybrids made up of the $+$ band and the $-$ band. 
For small enough $\mu$ and $\lambda$, the intensity switches between
the two in such a manner that we have one approximately
continuous basic mode and another much weaker shadow mode that
is also approximately continuous.\begin{figure}
\epsfxsize=4.5in
\centerline{\epsffile{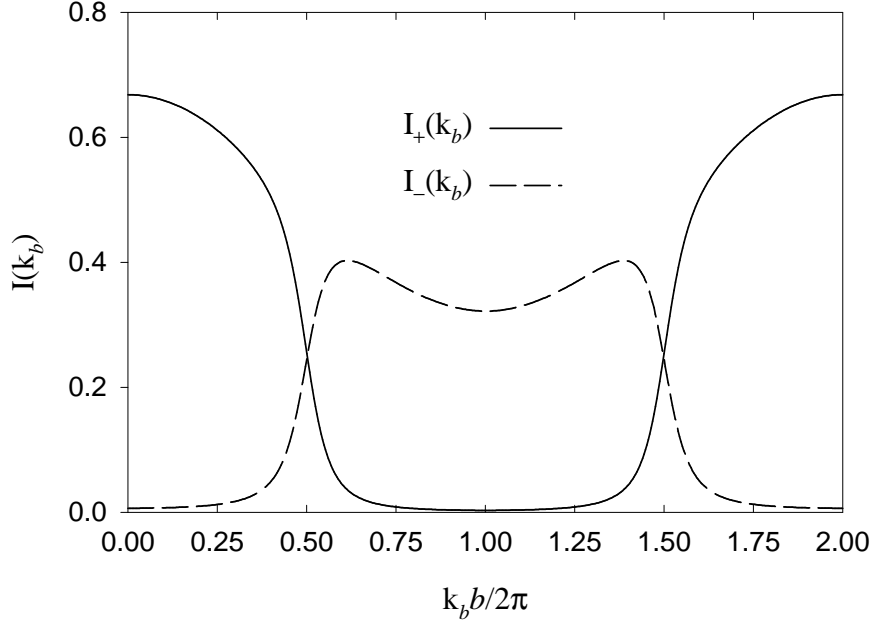}}
\vspace{0.1in}
\caption{Intensity at $k_a=0$ of the two single particle bands
when the interactions felt by the even and
odd dimers are different. For clarity, the $\sin^2(k_b w/2)$
modulation of the intensity is not included in the plots.
The values of $\lambda$, $\mu$ and $g$ are set as in Fig~{\protect
\ref{figmag1}}.}
\label{figmag2}
\end{figure}
These results are summarized in Fig~\ref{figmag1} and Fig~\ref{figmag2}.
Of course, the avoided level crossing between the two bands
leads to a small jump in position of both the basic and the
shadow mode situated near $\pi/b$ and $3\pi/b$. Thus the
intensity at the shadow positions and the gap introduced 
by the avoided level crossing are 
sensitive indicators of the difference between the
magnetic environments of even and odd dimers.
As mentioned earlier, all experiments to date~\cite{oldexp,ens,agthesis}
are consistent with the absence of extra modes at $k_a = 0$, but in
the absence of any straightforward symmetry reason,
more
experiments at $k_a = 0$ with better sensitivity and statistics are necessary
before one can completely rule out the existence
of such complications in the magnetic Hamiltonian of the
system.

\section{Conclusion}
The calculations presented here show that simple geometric effects 
can lead to the presence of the shadow bands observed recently in VOPO. 
These modes are similar to the `optic' modes that
can arise generally in coupled alternating-chain systems with more than 
one dimer per unit cell, for example the `chain' layers in 
Sr$_{14}$Cu$_{24}$O$_{41}$~\cite{matsuda}. However, in VOPO
the shadow modes arise from a single chain.
It should be possible to test our proposal by comparing the
experimentally observed
intensity ratios with our predictions.
Moreover, as mentioned earlier, the intensity ratios can
also distinguish between single-particle and bound state modes
The possible contribution of the triplet bound state
to the spin dynamics in VOPO remains an open question at this time.

\section{Acknowledgements}
KD would like to thank the Neutron Scattering Group at ORNL for hospitality,
and ORISE for financial support for a visit during which this work
was begun.
We thank T. Barnes, M. Enderle, B.C. Sales, H. Schwenk and D.A. Tennant
for numerous helpful discussions. Work at Princeton was supported
by NSF grant DMR-9809483.
The Oak Ridge National Laboratory is
managed for the U.S. D.O.E. by Lockheed Martin Energy Research 
Corporation under contract
DE-AC05-96OR22464.

\end{document}